\documentclass[aps,twocolumn,showpacs,floats,superscriptaddress]{revtex4}
\usepackage{graphicx}
\usepackage{dcolumn}
\usepackage{color}
\usepackage{bm}
\begin{document}


\author{Nikolay  Prokof'ev }
\author{Boris Svistunov}

\affiliation{Department of Physics, University of Massachusetts,
Amherst, MA 01003, USA} \affiliation{Russian Research Center
``Kurchatov Institute'', 123182 Moscow, Russia}

\title{Fermi-Polaron: Diagrammatic Monte Carlo for Divergent
Sign-Alternating Series}

\begin{abstract}
Diagrammatic Monte Carlo approach is applied to a problem of a
single spin-down fermion resonantly interacting (we consider
the universal limit of short-range interaction potential with
divergent s-wave scattering length) with the sea of
ideal spin-up fermions. On one hand, we develop a generic,
sign-problem tolerant, method of exact numerical solution of
polaron-type models. On the other hand, our solution is important
for understanding the phase diagram and properties of the BCS-BEC
crossover in the strongly imbalanced regime. This is the first,
and possibly characteristic, example of how the Monte Carlo
approach can be applied to a divergent sign-alternating
diagrammatic series.

\end{abstract}






\maketitle

In this Letter we introduce a novel technique for
studies of polaron type models which is equally
important from both physical and technical points of view.
On the physical side, we find a controllably-accurate numeric
solution for the problem of a single spin-down fermion resonantly
interacting with the sea of ideal spin-up fermions. This problem
naturally arises in studies of the  BCS-BEC crossover in the
strongly imbalanced regime \cite{Bulgac,Lobo}. In the case of
stability of the dilute spin-down sub-system in the ideal spin-up
fermi-sea, the solution of the single-particle problem would
naturally define the phase diagram in the vicinity of the
multicritical point (so-called M-point), discussed recently  by
Sachdev and Yang \cite{Sachdev}, where four different phases meat.
Even if the M-point is thermodynamically unstable---as is actually
suggested by the analysis of Giorgini and collaborators on the basis
of their fixed-node Monte Carlo simulations \cite{Giorgini}---the
single-particle M-point, corresponding to the critical interaction
strength at which the spin-down fermion forms a bound state with a
spin-up fermion thus becoming a spin-zero composite boson, is of
interest on its own. Also, for making a definitive theoretical
conclusion about thermodynamic (in)stability of the dilute spin-down
sub-system one still has to find an unbiased solution to the
single-particle problem.

On the technical side, we solve the Fermi-polaron problem by
diagrammatic Monte Carlo (MC) technique, which proved very
efficient for electron-phonon polaron problems \cite{polaron}.
Though the diagrammatic series for the Fermi-polaron are quite
similar, the crucial difference is that in the Fermi system we
have to deal with the sign-alternating, divergent (at least for
strong coupling) series. Under these conditions a direct summation
of all relevant Feynman diagrams for the Green's functions is not
possible, and one has to develop additional tools for (i) reducing
the number of diagrams by calculating self-energies rather than
Green's functions, and (ii) extrapolating MC results to the
infinite diagram order for a divergent series. We find that the
series for the Fermi-polaron are Riesz-summable and the
numerically exact solution within the diagrammatic MC approach
does exist. We believe that our findings are important in a much
broader context since the diagrammatic MC approach to the
many-body problem has essentially the same structure.

In what follows, we will be using the term `polaron' in a narrow
sense, i.e. only for the unbound fermionic spin-down excitation. For
the composite boson we will be using the term `molecule'. If polaron
is a well-defined elementary excitation, its energy $E({\bf p})$ can
be extracted from the pole in the single-particle spin-down Green's
function in the frequency-momentum representation, by solving the
equation
\begin{equation}
[G^{(0)}_\downarrow (\omega=E({\bf p}),\,  {\bf p})]^{-1}\, =\,
\Sigma_\downarrow (\omega=E({\bf p}),\,  {\bf p})\,\; ,
\label{pole}
\end{equation}
where $G^{(0)}_\downarrow$ is the vacuum Green's function for the
spin-down particle and $\Sigma_\downarrow$ is the self-energy of
spin-down particle in the Fermi-sea of spin-up particles. In
diagrammatic Monte Carlo we traditionally use the
imaginary-time--momentum representation to circumvent the problem
of dealing with the singular structure of free propagators and
alleviate the sign problem. Remarkably, in order to find the
polaron energy we do not have to perform a numeric analytic
continuation from imaginary to real frequencies, because in the
imaginary-time--momentum representation Eq.~(\ref{pole})
translates into
\begin{equation}
[G^{(0)}_\downarrow (\omega=E({\bf p}),\,  {\bf p})]^{-1}\, \,=
\int_0^\infty \Sigma_\downarrow (\tau,\, {\bf p}){\rm e}^{E({\bf
p})\tau}d\tau  \; . \label{pole2}
\end{equation}
The equivalence of Eqs.~(\ref{pole}) and (\ref{pole2}) readily
follows from the fact that for the spin-down particle there is a
freedom of choosing the `chemical potential' $\mu_\downarrow$ (for a
single particle this is just a uniform external potential which does
not affect its physical properties). This freedom can be utilized
for fine-tuning $\mu_\downarrow = E({\bf p})$, in which case the
solution of Eq.~(\ref{pole}) corresponds to zero frequency. Then, by
noting that zero real frequency is the same as zero imaginary
frequency, and utilizing the trivial dependence of self-energy on
$\mu_\downarrow$, namely, $\Sigma_\downarrow (\tau,\, {\bf p}, \,
\mu_\downarrow)\, =\, \Sigma_\downarrow (\tau,\, {\bf p}) {\rm
e}^{\mu_\downarrow \tau}$, one proves Eq.~(\ref{pole2}).

\begin{figure}[t]
\vspace*{-0cm}
\includegraphics[bb=320 390 535 430, width=\columnwidth]{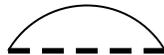}
 \caption{ The first-order diagram for $\Sigma_\downarrow$,
 consisting of the $T$-matrix propagator (heavy dashed line) and
 spin-up propagator (solid arc). } \label{fig1}
\end{figure}

We obtain self-energy $\Sigma_\downarrow (\tau,\, {\bf p})$ by
means of diagrammatic MC which works with standard diagrammatic
expansions for Green's functions by interpreting them as a
statistical ensemble. In our case, the diagrams are constructed
from the following lines: (i) the already-mentioned vacuum
propagator $G^{(0)}_\downarrow (\tau ,p) =e^{(\mu_\downarrow -
p^2/2m)\tau }$ where $m$ is the particle mass; it will be
represented by thin horizontal straight lines, (ii) the spin-up
propagators $G_\uparrow (\tau ,p>k_F)=e^{(\epsilon_F - p^2/2m)\tau
}$ and $G_\uparrow (\tau, p<k_F)=e^{-(\epsilon_F - p^2/2m)\tau }$
(where $k_F$ and $\epsilon_F$ are the Fermi momentum and Fermi
energy of spin-up particles); they are depicted with thin-line
arcs, and (iii) the $T$-matrix propagator $\Gamma(\tau ,p)$. The
specifics of the resonant problem is that $T$-matrix has to be
considered as a separate diagrammatic element which sums ladder
diagrams for the short-range pair interaction potential. This
summation takes the ultra-violet physics into account exactly and
allows to express $\Gamma(\tau ,p)$ in terms of the $s$-scattering
length $a$. The ladder structure of diagrams absorbed in the
$T$-matrix explains why we treat it as a `pair propagator' and
depict it with a (dashed heavy) line. The exact expression for
$\Gamma(\tau , {\bf p})$ in the zero-range limit in the
frequency-momentum representation reads
\begin{equation}
\Gamma^{-1}(\omega,p)\, =\,{m\over 4\pi a}\, -\, {m\over 8\pi
}\sqrt{p^2-4m\eta}-\Pi(p,\eta)
 \; , \label{Gamma}
\end{equation}
\begin{equation}
\Pi(p,\eta)\, =\, \int_{q\leq k_F} { d{\bf q} /(2\pi)^3 \over
q^2/2m+({\bf p}-{\bf q})^2/2m -\eta  }\; . \label{a}
\end{equation}
\begin{equation}
\eta\, =\, \omega + \varepsilon_F \; . \label{b}
\end{equation}
As an illustration, in Fig.~\ref{fig1} we present the first-order
diagram for $\Sigma_\downarrow$.

\begin{figure}[t]
\vspace*{-0cm}
\includegraphics[bb=250 390 600 450, width=\columnwidth]{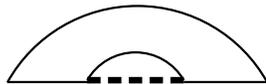}
 \caption{ The first diagram for molecular self-energy $K$,
 consisting of one $T$-matrix propagator (heavy dashed line), two
 spin-up propagators (solid arcs), and two spin-down propagators (solid straight lines). }
 \label{fig2}
\end{figure}

The energy of the molecule, $E_m({\bf p})$, cannot be extracted
from Eqs.~(\ref{pole}), (\ref{pole2}), because the corresponding
pole exists only in the two-particle (spin-up + spin-down) or
``pair" channel. Due to the fact that the resonant $T$-matrix is
formally playing the role of a pair propagator, the theory of the
molecular pole is analogous to the Dyson-equation theory of the
polaron pole. Diagrammatically, it arises from the summation of
geometric series of products $\Gamma K \Gamma K \Gamma K \ldots$
where $K$ is the $\Gamma$-irreducible diagram in the two-particle
channel---a direct analog of self-energy of the single-particle
channel (see Fig.~\ref{fig2} for an illustration).
Correspondingly, the molecular energy is found by solving an
equation
\begin{equation}
\Gamma^{-1} (\omega=E_m({\bf p}),\,  {\bf p})\, = \, \int_0^\infty
K (\tau,\, {\bf p})\, {\rm e}^{E_m({\bf p})\tau}d\tau \; .
\label{polem}
\end{equation}

We  found neither analytical expression for $\Gamma(\tau, p)$, nor
fast way of tabulating it with high accuracy using inverse Laplace
transform of the frequency-momentum representation
Eq.~(\ref{Gamma}). Instead, we applied recently-developed bold
diagrammatic Monte Carlo (BMC) technique \cite{BMC} to obtain
$\Gamma(\tau, p)$ numerically, by relating it to the vacuum
$T$-matrix the analytic expression for which in $(\tau,
p)$-representation is readily obtained. This was the first
practical application of BMC, see Ref.~\cite{BMC} for details.

The updates we use for sampling the diagrammatic series do not
differ dramatically from the ones described in  literature, and we
leave the corresponding discussion to a full-sized paper
\cite{paper}. Instead, we concentrate on the convergence issues.
In contrast to previous examples of diagrammatic Monte Carlo, the
series for the resonant Fermi-polaron problem turns out to be
divergent. However, the Ces\`aro-Riesz summation techniques fix
the problem. For the quantity of interest---polaron or molecule
self-energy---we construct the Ces\`aro-Riesz partial sums,
$\Sigma (N_*)= \sum_{N=1}^{N_*} D_N F_N^{(N_*)}$, defined as sums
of all diagrams up to order $N_*$ [the diagram order, $N$, is
defined by the number of spin-up propagators] with the $N$-th
order terms being multiplied by the factor
\begin{equation}
F_N^{(N_*)} = [(N_*-N+1)/N_*]^\delta \;, ~~~~~
\mbox{(Ces\`aro-Riesz)}\;. \label{factor1}
\end{equation}
Here $\delta > 0$ is a fixed parameter ($\delta = 1$ corresponds
to the Ces\`aro method). If the series is Riesz-summable, then the
answer in the  $N_* \to \infty$ limit does not depend on $\delta$.
The freedom of choosing the value of Riesz's exponent $\delta$ is
used to optimize the convergence.
\begin{figure}[t]
\vspace*{-0cm}
\includegraphics[bb=0 15 270 210, width=\columnwidth]{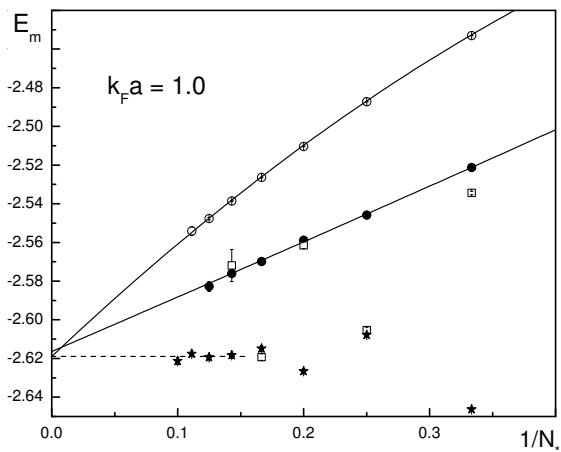}
 \caption{ The molecule  energy (at $k_F\, a=1$) as a function
of truncation parameter $N_*$ for different summation techniques:
Ces\`aro (open squares), Riesz $\delta=2$ (filled circles, fitted
with the parabola $y=-2.6164 + 0.28013x + 0.01638 x^2$), Riesz
$\delta=4$ (open circles, fitted with the parabola $y=-2.6190 +
0.61635 x - 0.3515 x^2$), advanced summation described in the text
(stars). In the $\delta=2$ case, the odd-even oscillations are
strongly suppressed, as compared to Ces\`aro case, but still visible
at higher resolution.  } \label{fig3}
\end{figure}
\begin{figure}[t]
\vspace*{-0cm}
\includegraphics[bb=0 20 290 230, width=\columnwidth]{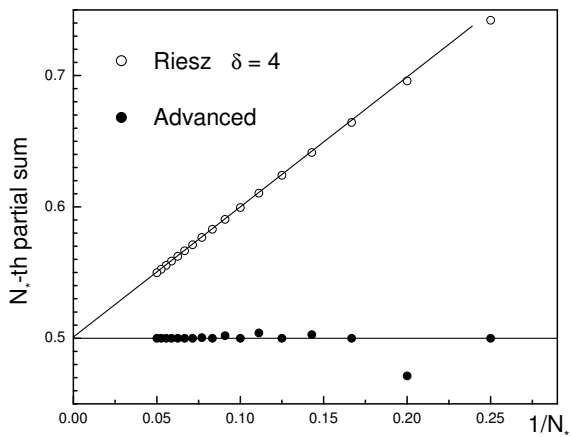}
 \caption{ Summing the Grandi series, $1-1+1-1+\ldots = 1/2$, by Riesz method and by the
advanced method described in the text. The solid lines are to
guide an eye.} \label{fig4}
\end{figure}
\begin{figure}[t]
\vspace*{-0cm}
\includegraphics[bb=5 15 280 210, width=\columnwidth]{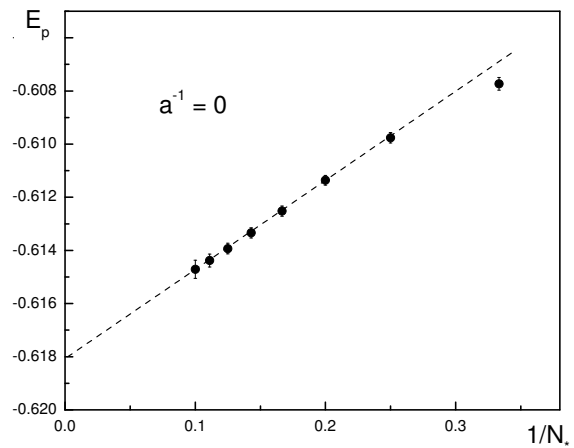}
 \caption{ The polaron energy (at the unitarity point $a^{-1}=0$) as a function
 of truncation parameter $N_*$ for the advance summation
 technique. The asymptotic behavior at $1/N_* \to 0$ is perfectly
fit by a straight line.} \label{fig5}
\end{figure}

For the $N_*$-truncated and reweighed series we first determine
the polaron and molecule energies and then study their dependence
on $N_*$ as $N_* \to \infty$. Figure~\ref{fig3} illustrates the
procedure. The odd/even oscillations are very pronounced for
$\delta=1$ hinting at the absence of convergence of the original
series but are strongly suppressed for larger values of $\delta$.
With $\delta=4$ we were not able to resolve odd-even oscillations,
but the smoothness of the curve here comes at the expense of
increased curvature, which renders the extrapolation to the $1/N_*
\to 0$ limit vulnerable to systematic errors. Empirically, we
constructed a sort of generalized Ces\`aro-Riesz factor
$F_N^{(N_*)}$ which leads to a much faster convergence (cf.
Fig.~\ref{fig4}); in what follows we refer to it as an {\it
advanced} summation technique
\begin{equation}
F_N^{(N_*)}\, =\,
C_{N_*}\sum_{m=N}^{N_*}\exp{\left[-{(N_*+1)^2\over m(N_*-m+1)}
\right] } \; , \label{factor2}
\end{equation}
\begin{equation}
C_{N_*}^{-1}\, =\, \sum_{m=1}^{N_*}\exp{\left[-{(N_*+1)^2\over
m(N_*-m+1)} \right] } \; . \label{c}
\end{equation}
For the molecular energy, the advanced summation demonstrates no
visible linear slope in the $1/N_* \to 0$ limit. For the polaron
energy, there is a small linear slope, but without visible
curvature, see Fig.~\ref{fig5}.

To ensure that both the code and data processing are error free, we
check our answer for the molecule groundstate energy against the
asymptotic solution
\begin{equation}
E_m\, = \, -{1\over ma^2}\, -\, \varepsilon_F\, +\,  {2\pi
\tilde{a} \over (2/3)m } n_\uparrow ~~~~~~(k_Fa\, \to 0\, ) \, ,
\label{mol_asym}
\end{equation}
in which the first term in the r.h.s. is the binding energy of
molecule in vacuum, the second term reflects finite chemical
potential for spin-up fermions, and the third term comes from the
mean-field interaction between the compact molecule and the
remaining spin-up particles. Correspondingly, $\tilde{a} \approx
1.18 a$ is the molecule-fermion $s$-scattering length
\cite{Skornyakov}. The value of $\tilde{a}$ is based on the
non-perturbative solution of the three-body problem, and thus
provides a robust test to the entire numerical procedure of
sampling divergent sign-alternating diagrammatic series. Our data
are in a perfect agreement with the $\tilde{a}\approx 1.18a$
result within the statistical uncertainty of the order of $5\%$.

\begin{figure}[t]
\vspace*{-0cm}
\includegraphics[bb=10 5 280 220, width=\columnwidth]{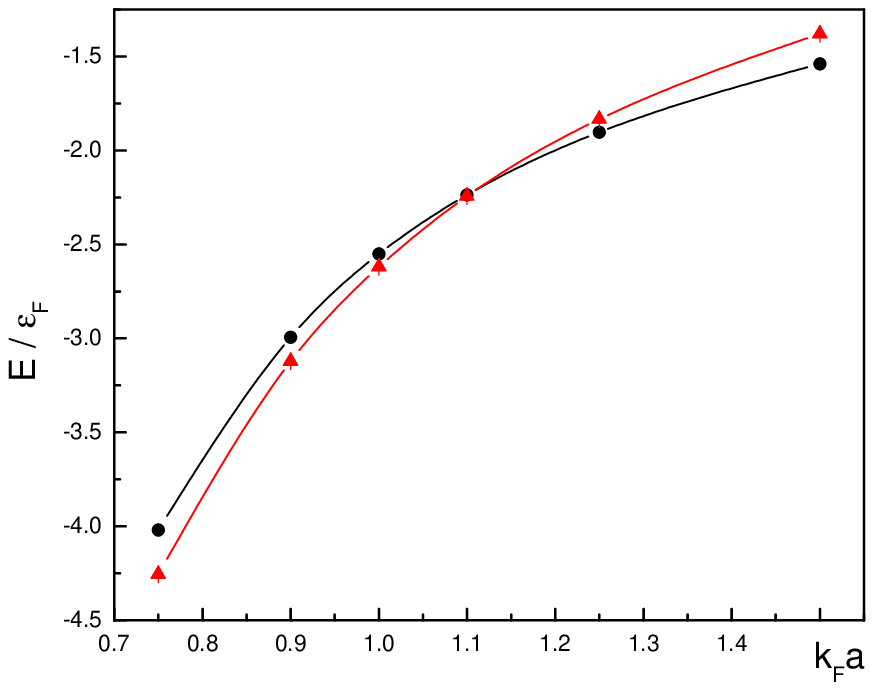}
\includegraphics[bb=10 15 280 210, width=\columnwidth]{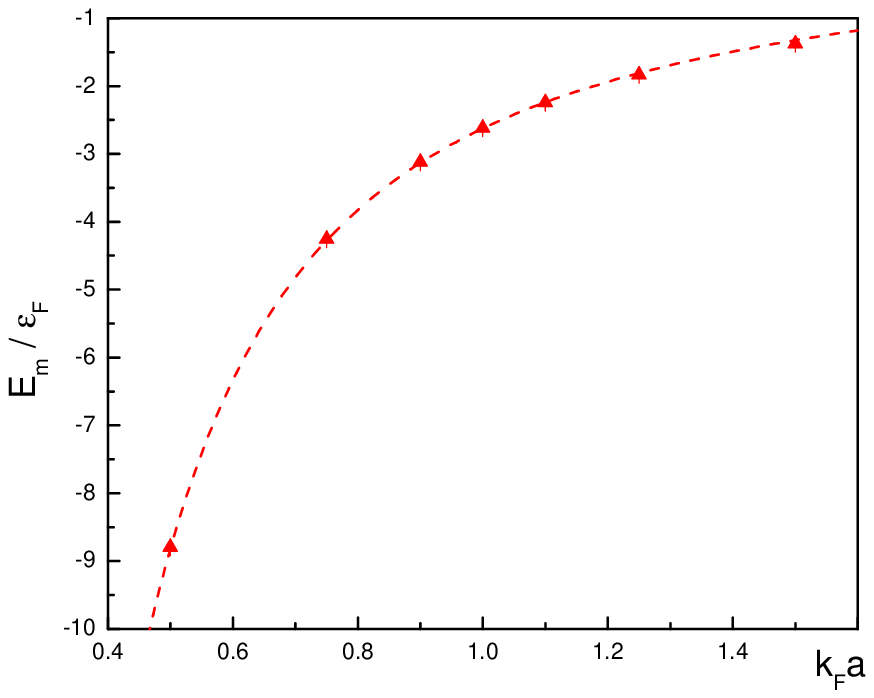}
 \caption{ Polaron (circles) and molecule (triangles) energies
 (in units of $\varepsilon_F$) as functions of $k_F a$. The dashed line on the lower panel
 corresponds to Eq.~(\ref{mol_asym}). }
 \label{fig6}
\end{figure}
\begin{figure}[t]
\vspace*{-0cm}
\includegraphics[bb=5 15 280 240, width=\columnwidth]{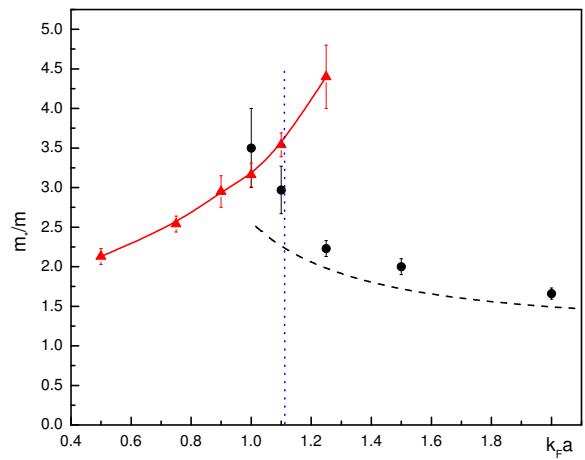}
 \caption{ Polaron (circles) and molecule (triangles) effective
 mass  (in units of $m$) as functions of $k_F a$.
 The vertical dotted line stands for $(k_F a)_c=1.11$.
 The dashed line  is the contribution from the first diagram
 \cite{Combescot}.} \label{fig7}
\end{figure}

In Fig.~\ref{fig6}, we show polaron and molecule energies in the
region of $k_Fa\sim 1$ where the nature of the quasiparticle state
changes. The $M$-point is found to be at $(k_Fa)_c = 1.11(2)$. The
values of both polaron and molecule energies are in excellent
agreement with the fixed-note Monte Carlo simulations
\cite{Lobo,Giorgini}. Interestingly, the polaron self-energy is
nearly exhausted by the first-order diagram considered in
Ref.~\cite{Combescot}, see also Fig.~\ref{fig5}.

A comment is in order here on why the polaron (molecule) remains a
good elementary excitation at $k_Fa < (k_Fa)_c$ ($k_Fa > (k_Fa)_c$)
and the $M$-point is an exact crossing. In the limit of $k_Fa \to
(k_Fa)_c$, this is guaranteed by the phase-space argument. The
conservation of energy, momentum, and spin projection dictate that
the leading decay channel involves four quasiparticles in the final
state: the polaron decays into molecule, two holes and one spin-up
particle; the molecule decays into a polaron, one hole, and two
spin-up particles. Correspondingly, the decay width---proportional
to the four-particle phase-space volume allowed by the conservation
laws---gets negligibly small as compared to the energy difference
$|E_m-E_p|$ at $k_Fa \to (k_Fa)_c$. Our numeric results show that
$|E_p-E_m| \ll |E_p|, |E_m|$ even when $k_Fa$ is substantially
different from $(k_Fa)_c$. Since the phase-space volume is
determined by $|E_p-E_m|$, this explains why the width of the
decaying quasiparticle remains small and goes beyond our numeric
resolution.

The data for the effective mass is presented in Fig.~(\ref{fig7}).
As expected, at the crossing point the effective mass curve is
discontinuous. Note that good agreement with Eq.~(\ref{mol_asym})
for $E_m$ at all couplings up to the $M$-point is slightly
misleading since Eq.~(\ref{mol_asym}) assumes that molecules are
compact and their mass is $2m$. The actual effective mass is
significantly enhanced in the vicinity of the $M$-point.

Summarizing, the problem of resonant Fermi-polaron has been solved
by diagrammatic MC. In particular, the point where the groundstate
switches from the single-particle (fermionic) sector to
two-particle (bosonic) sector is found to be at $k_Fa = 1.11(2)$.
On the computational side, it has been discovered that while the
diagrammatic series is divergent, the fermionic sign-alternation
of the diagrams renders the series summable by the
Ces\`aro-Riesz-type methods and suitable for the diagrammatic MC
routine. Whether this {\it sign blessing} is a specific feature of
the problem solved here, or a generic feature of fermionic and,
possibly, other diagrammatic expansions, is a major question which
we will address in future.

We are grateful to C. Lobo, S. Giorgini, and R. Combescot for
valuable discussions and data exchange. The work was supported by
the National Science Foundation under Grants PHY-0426881 and
PHY-0653183. We also acknowledge a support from KITP, Santa
Barbara (under NSF Grant PHY05-51164) and PITP, Vancouver.

\end{document}